\documentclass{ringb99}
\usepackage{graphics}
\begin{document}
\title{	Some Aspects of Inverse Compton Emission from 3K Background Photons}
\author{D. E. Harris\inst{1} \and L. S. Moore\inst{2}}  
\institute{Smithsonian Astrophysical Observatory, 60 Garden St.,
Cambridge, MA 02138, USA; harris@cfa.harvard.edu \and Bryn Mawr
College, Bryn Mawr, PA  USA; lmoore@brynmawr.edu}
\maketitle

\begin{abstract}

We review the history and a few features of IC/3K emission.  We 
discuss observing strategies, the physical parameters which
can be derived directly or indirectly from successful detections, and
the disparities between magnetic field strength estimates found by
different methods.
 
\end{abstract}

\section{Introduction}

The recent history of inverse Compton emission from cosmic
background photons and relativistic electrons (IC/3K) began during the
period when the origin of extended X-ray emission from clusters of
galaxies was unclear.  Felten and Morrison (1966), Blumenthal \& Gould
(1970), and Jones, O'Dell \& Stein (1974) were among those who worked
out the details of IC/3K, but not until iron lines were detected in
the X-ray spectra was it generally accepted that the bulk of the X-ray
emission was caused by the thermal bremsstrahlung process.

Although there were many attempts to isolate IC/3K emission from
clusters, these were thwarted by the strong thermal emission which is
quasi ubiquitous in clusters.  The recent breakthrough came from the
dual detections of IC/3K from the lobes of Fornax A: spatially with
the ROSAT PSPC (Feigelson et al. 1995) and spectrally with ASCA (Kaneda
et. al. 1995).  In the last two years, there have been papers
published on other sources which are relatively free of hot gas
emission and from cluster radio halos, both from EUV excesses and from
the harder spectral component at energies for which the thermal X-rays
are dropping exponentially.
                                                                          
\section{Salient Features}

In this section, we list a few of the attributes of IC/3K emission
which are often overlooked.

\subsection{Every non-thermal radio source also emits IC/3K emission}

IC/3K emission is mandatory: every (non-thermal) radio source in
the universe is an emitter.  This statement entails only the assumption 
that the 3K background is indeed universal.  Thus the only problem is
to see if it is strong enough for detection with a given system, and if
it can be separated from other emissions.

\subsection{Electron Energy and the observing frequency}

At a given (X-ray) energy, we will be sampling the relativistic
electron spectrum at the same value of electron energy ($\gamma$) for
every source in the universe.  This is because the peak frequency,
$\nu_o$, of the cosmic background spectrum (where the major
contribution to the photon energy density occurs) increases as (1+z).
A given electron with Lorentz energy, $\gamma$, emits most of its
radiation at $\gamma^2~\times~\nu_o$ and the emitted photons at that
energy are redshifted by (1+z) when observed on Earth.  Thus soft
X-rays (1-2 keV) sample $\gamma~\approx$~1000 no matter what the
redshift.  Attempts to integrate these electrons in order to explain a
significant part of the soft X-ray background have not been successful.

\subsection{The importance of determining the amplitude of the
electron spectrum at low energies}

For B$<$5$\mu$G, low energy electrons (e.g. $\gamma<$1000) are not
visible via their synchrotron (radio) emission because the ionosphere
blocks frequencies below $\approx$~20 MHz.  Since these electrons
normally have long lifetimes in the weak magnetic fields typically
found in extended radio sources, they accumulate over the life of a
radio source.  In a sense we are sampling an encapsulated history of
the radio source: the electron spectrum at low energies tells us about
all the electrons produced over the source's lifetime.

\section{Observing Strategies - avoid hot gas!}
 
\begin{itemize}

      \item By going to higher energy: e.g. BeppoSax and RXTE

      \item By getting outside clusters: e.g. Fornax A lobes and relic
radio galaxies such as 0917+75 (Harris et. al. 1995).

      \item Choose radio sources with large numbers of low energy
electrons.  This can be achieved by choosing sources likely to have 
relatively more low energy electrons and to have weak magnetic fields
since these require more electrons to get a given radio intensity.
Therefore, steep spectrum, low brightness radio sources are best.

\end{itemize}

\section{Rewards}

Although most of us like to believe that a successful detection of
IC/3K yields the average magnetic field strength directly, what we really
measure is the amplitude of the electron spectrum at some low energy.
To obtain the average field strength, there is still the uncertainty
of the form of the electron spectrum between the direct measurement
afforded by the IC/3K observation and the segment of the electron
spectrum responsible for the radio emission.  This is illustrated in
figure 1.  Since we can't be sure that a single power law extrapolates
from the radio derived segment of the spectrum to the X-ray derived
amplitude, the average value of B is uncertain (even if
the ROSAT point were not an upper limit).

\begin{figure}
\resizebox{8.5cm}{10.0cm}{\includegraphics{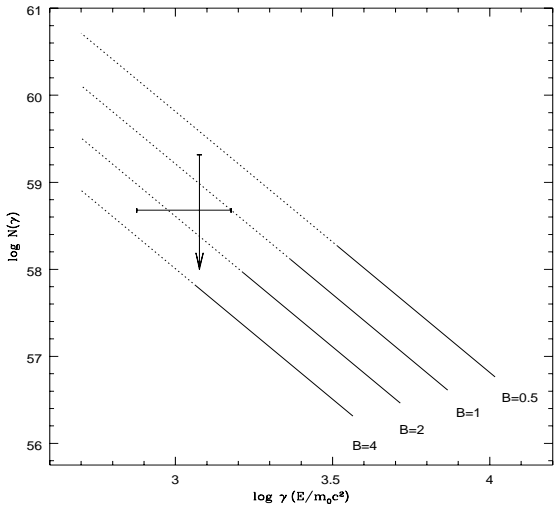}}
\vskip -1.8cm
\caption[Relativistic Electron Spectrum for radio relic
0917+75]{Constraints on the electron spectrum for the relic radio
galaxy, 0917+75 (taken from Harris et al. 1995).  Heavy lines show
possible segments of the electron spectrum, each labelled with a trial
value of the magnetic field strength (in $\mu$G).  The ROSAT derived
upper limit is the only firm datum.}

\end{figure}

The electron spectrum below $\gamma$=1000 is of particular interest
because in weak field regions (and z$<$0.5), E$^2$ halflives can
approach 10$^9$~yr.  Therefore, the total number of electrons in this
energy range serves as a diagnostic of the total energy of a source
during its life.  While soft X-rays (e.g. 1.5 keV) provide a direct
measurement of the number of electrons at $\gamma$=1000, EUVE data can
provide estimates at $\gamma$=300.  In Table 1 we have calculated
log~N (100$<\gamma<$1000) for a variety of radio sources by
extrapolating their observed spectra to lower energies.

\small
\textbf{Table 1  Number of low energy electrons from extrapolated
radio spectra}
\tiny

\begin{center}
\begin{tabular}{lllllccc}

Source	&Component&	z	&$\alpha_r$	&B(eq)	&logV	&logN	&$\tau$ \\	
&&&&($\mu$G)	&(cm$^{3}$)&&		(yrs)
\\ \hline
&&&&&&& \\								
FRII		&&&&&&& \\						
Cyg A	&2 lobes&0.057	&0.6	&25	&69.83	&63.59	&1E7 \\	
&&&&&&& \\								
RELICS&&&&&&& \\						
0917	&total&	0.12&	1.0&	0.7&	72.77&	64.23&	8E8 \\	
Coma&	total&&		1.18&	0.3&	72.19&	64.53&	1E9 \\	
CenB&	1 lobe&	0.012&	0.8&	1.5&	71.14&	63.32&	9E8 \\	
1358+30&&	0.11&	0.72&	0.9&	72.00&	63.22&	8E8 \\	
1401-33&&	0.0136&	1.44&	8&	71.67&	64.91&	1E9 \\	
3C 326&	giantRG&	0.089&	0.82&	0.8&	73.46&	64.75&	9E8 \\	
IC2476&&		0.027&	1.1&	0.7&	71.57&	63.80&	1E9 \\	
&&&&&&& \\						
FRI&&&&&&& \\						
3C 31	&&0.016&	0.63&	1.5&	70.61&	61.97&	9E8 \\	
0715&	1arm&	0.07&	1.1&	 4&	...&	62.40&	... \\	
1718&	halo&	0.162&	1.3&	 6&	...&	62.90&	... \\	
&&&&&&& \\								
HALOS&&&&&&& \\				
Coma&	total&	0.023&	1.34&	0.5&	72.64&	65.28&	1E9 \\	
% &&&&&&& \\								
% \multicolumn{2}{l}{UNSPECIFIED}&&&&&& \\
% 3C464&&		0.055&	1.9&	 7&	69.11&	64.72&	2E8 \\
\end{tabular}
\end{center}

\small
\vskip 0.2cm  
\noindent
{\bf Notes to Table 1}

\noindent
z is the redshift\\
$\alpha_r$ is the radio spectral index\\
B(eq) is the equipartition field\\
V is the source volume\\
N is the integrated number of electrons for 100$<\gamma<$1000\\
$\tau$ is the halflife for $\gamma$=1000 electrons.

\normalsize

\vskip 0.2cm

While there are considerable uncertainties on the magnetic field and
volume estimates, a substantial difference is seen for log N amongst
Cyg A and relics; FRIs; and the Coma halo.  These values support the
notion that the relics were once FR II radio galaxies and that if
cluster halos arise from the remnants of FR I galaxies such as tailed
radio galaxies, it would require of order 1000 such contributions in
10$^9$~years.

Once an estimate of the average magnetic field strength is obtained,
inferences can be made on the total energy density, $u(tot)$, and thus
on the composition of the relativistic particles.  There are two
contributions to the particle energy density, $u(p)$, which are not
'counted' by the radio observations: low energy electrons and
relativistic protons.  For low brightness radio sources, it is
probably the case that most of the source is 'relaxed' in the sense
that there are not active shock regions (e.g. hotspots) and
equipartition between magnetic field energy density, $u(B)$, and
$u(p)$ is a reasonable expectation.  Thus knowledge of $u(B)$
translates to an estimate of $u(p)$ and $u(tot)$.  If the required
$u(p)$ is large enough, the presence of relativistic protons may be
indicated.  The major uncertainty in this process is probably the
unknown filling factor, $\phi$, and if $\phi$ is substantially less
than unity, the additional question: "Do the particles and field
occupy the same volume?".

\section{Discussion}

\subsection{Distinction between radio lobes (including relics) and cluster halos}

While we have some evidence that relativistic and thermal plasmas are
spatially distinct in the case of lobes of radio galaxies (e.g. Cygnus
A, Carilli et al. 1994), we do not have any confidence as to the situation
for cluster halos.  This is partly caused by the uncertainty
concerning the genesis of radio halos in clusters.  If relativistic
plasma is completely mixed with the thermal plasma, then it is
reasonable to assign a value close to unity for the filling factor, $\phi$,
but if the actual synchrotron emitting regions are to be distinct from
the thermal plasma (in the sense of a magnetic boundary separating the
two), then $\phi$ is most likely quite small.  In the former case
we should probably include the thermal energy density in any
application of the equipartition condition, whereas in the latter case,
we would be justified in balancing the relativistic particle energy 
density with the magnetic field energy density.  

In a sense, we have two distinct problems.  For relics and radio lobes
such as Fornax A, we do not have to contend with excessive thermal emission
and can get a fairly clean measurement if IC/3K intensity and distribution.
But for clusters, it is only with difficulty that we avoid the thermal
contamination.  

What are the implications of the fact that some estimates of the Coma
field strength are in reasonable accord with the minimum energy
B field for $\phi$=1 and $u(p)$=$u(B)$?  If the synchrotron emitting
plasma were distinct from the cluster gas and $\phi<<$1, then the
equipartition field B(eq) would be much larger than the IC/3K
estimate, and that does not seem to be the case.

\subsection{Current IC/3K detections and the implied B fields}
	
Using the relationships of Harris \& Grindlay (1979), we have
calculated the value of the average magnetic field strengths in some
of the sources for which IC/3K detections have been claimed or an
upper limit has been measured.  These are presented in Table 2
together with magnetic field estimates published by the authors.

\subsection{Comparison of B field estimates}

The B field estimates for clusters discussed at this workshop can be
roughly divided into 3 categories, depending on the method used and
the type of source.

\noindent
\textbf{B=0.1 to 0.2 $\mu$G}\\
These sorts of values were put forward on the basis of the excess
X-ray flux observed over that expected from the hot gas in the Coma
cluster.  Both the EUV excess and the BeppoSax measurements yield B
values $<$0.2 $\mu$G.  Fusco-Femiano et al. (1999) give
fx(20-80keV)=2.2$\times$10$^{-11}$ ergs cm$^{-2}$ s$^{-1}$ and
B$\approx$ 0.15~$\mu$G.  Note however that Henriksen (1998) using ASCA
and HEAO-I data obtained an upper limit of
fx(20-60keV)$<$2.9$\times$10$^{-12}$ ergs cm$^{-2}$ s$^{-1}$ and
B$>$0.26~$\mu$G.

\noindent
\textbf{B$\approx$~few $\mu$G}\\
Magnetic field strengths of a few $\mu$G have been derived from
equipartition calculations for the Coma radio halo, from the IC/3K
estimates for the lobes of Fornax A, and from some
estimates for clusters from Faraday rotation.

\noindent
\textbf{B = 5 to 20 $\mu$G}\\
These stronger B field estimates generally come from the Faraday
rotation measurements of unresolved sources in or behind clusters.
The greatest uncertainty here is the scale size of magnetic field
cells (for which the fields can reverse direction).  If there are a
large number of cells along the line of sight which have sufficient
field strength and electron density to make a substantial contribution
to the Faraday rotation, then the required B field will be
significantly larger than for the case of only a few cells.

While we are not in a position to resolve the disparity in these B field
estimates, we suspect that typical field strengths in clusters and
relaxed radio lobes probably lie within the range 0.5$<B<$3$\mu$G.  If
this were to be the case, then published EUV and hard X-ray excesses
are either wrong or arise from an emission process other than IC/3K
and clusters have large scale coherent fields so that there are
relatively few field reversals along the line of sight.

\section{Conclusion}

We have a fairly good idea of the problems involved in using IC/3K
emission to derive physical quantities.  Most of these difficulties
are not going to go away soon, but the quality of the X-ray data
should improve significantly with the advent of several missions with
larger collecting area, better resolution, and extended frequency
coverage.

% \begin{acknowledgements}
% \end{acknowledgements}

\onecolumn
\normalsize
\begin{center}
\textbf{Table 2  Magnetic field estimates from IC/3K detections}
\begin{tabular}{llrrrrrrr}
&&\multicolumn{3}{c}{Luminosities}&\multicolumn{2}{c}{B
Fields(A)}&\multicolumn{2}{c}{B Fields(RX)} \\ \cline{3-5} \cline{6-7}
\cline{8-9}
&&Log Lr&Log Lx&Lr/Lx&B(eq)&B(ic)&B(eq)&B(ic) \\
&&\multicolumn{2}{c}{(erg/s)}&&\multicolumn{2}{c}{($\mu$G)}&\multicolumn{2}{c}{($\mu$G)}
\\
\multicolumn{2}{l}{CLUSTERS}&&&&&&& \\	                                                     
Coma 1	&	&40.85   &43.71   &0.001    &...    &0.15    &0.2
&0.05 \\  
&&&&&&&& \\
\multicolumn{2}{l}{RADIO LOBES}&&&&&&& \\
\multicolumn{2}{l}{Fornax A}&&&&&&& \\ 
&	Keast	&41.40   &40.95   &3.24     &...    &2.4     &1.0
&1.8 \\   
	&Kwest	&41.64   &40.89   &5.62     &...    &3.5     &1.0    &2.6   \\
	&Fwest	&41.78   &41.08   &5.01     &3.0    &1.9     &1.4    &2.0   \\
&&&&&&&& \\
\multicolumn{2}{l}{RELICS}&&&&&&& \\
A 85  & 0038-096	&41.80	&42.95	&0.07	 &...	&0.9	&2.2
&0.4 \\   
Cen B &$<$lobe$>$&	42.20   &41.50   &5.01   &  ...   & 3.1  &
1.4 &   1.7 \\
0917	&total	&41.83  &\llap{$<$}42.30  &\llap{$>$}0.34&     0.7&  
\llap{$>$}0.7     &0.7   &\llap{$>$}0.7   \\
\end{tabular}
\end{center}

\small

{\bf Notes to Table 2} \\ 

Luminosities are indicative; the radio band was generally 10$^7$ to
10$^{10}$~Hz (except as noted below).  The X-ray bands depend on the
telescopes used.  B(eq) is the equipartition field (filling factor of
unity and no protons) and B(ic) is the field strength derived from the
IC/3K measurements.  Both the original authors' values (A) and those
calculated from the equations in Harris and Grindlay (1979) ['RX'] are
given.

\vskip 0.2cm  

\noindent COMA 1 

Entries are based on the composite radio spectrum presented by
Fusco-Femiano et al. (1999); 'FF' hereafter.  The general properties
of the radio spectrum of the halo were checked against earlier
presentations.  Note however that whereas earlier work found
$\alpha_r$ around 1.2 to 1.3, FF argue for curvature above 100 MHz,
thus deducing a spectral break with an injection spectrum
characterized by $\alpha$=0.96.  For RX inputs, we take the
synchrotron band to be 30-630MHz, with a fiducial flux density of
S(100MHz)=10 Jy (from their figure).  The X-ray data are from BeppoSax
(ibid.).  FF combine two detector results and make a spectral fit with
two components (thermal and a power law).  For the power law, they get
$\alpha_x$ =0.6$\pm$0.4.  For the volume, we take the diameter used by
FF: 25' (1 Mpc).

\vskip 0.2cm  

\noindent Fornax A

We used two papers: Kaneda et al. (1995) - ASCA spectral analysis ("Keast" and
"Kwest"), and Feigelson et al. (1995)- spatial analysis from the ROSAT PSPC.
Feigelson only did the West lobe ("Fwest").

\vskip 0.2cm  

\noindent A85

The results are based on Bagchi et al. (1998) from a wavelet analysis of
ROSAT PSPC data.

\vskip 0.2cm  

\noindent CENTAURUS B (a.k.a. PKS1343-601)

This is a relic radio galaxy, somewhat obscured by the Milky Way.
Radio data are from McAdam (1991) and the X-ray data are from Tashiro
et al. (1998).  Since the X-ray data are for the entire source, and
since the lobes are not the same size and brightness, we model an
'average lobe', taking the size, r=300" (that of the S lobe), and
divide both radio and X-ray fluxes by 2.

\normalsize


\begin{thebibliography}{}

\bibitem{BPL98} Bagchi, J., Pislar, V., and Lima Neto, G.B. 1998 MNRAS
296, L23

\bibitem{BG70} Blumenthal, G.R. and Gould, R.J. 1970
Rev. Mod. Phys. 42, 237

\bibitem{CPH94} Carilli, C.L., Perley, R.A., and Harris, D.E. 1994
MNRAS 270, 173

\bibitem{FLKF95} Feigelson, E.D., Laurent-Muehleisen, S.A., Kollgaard,
R.I. Fomalont, E.B. 1995 ApJ 449, L149

\bibitem{FM66} Felten, J.E. and Morrison, P. 1966 ApJ 146, 686 

\bibitem{FDFGGMMS99} Fusco-Femiano, R., Dal Fiume, D., Feretti, L.,
Giovannini, G., Grandi, P., Matt, G., Molendi, S., and Santangelo,
A. 1999 ApJ 513, L21

\bibitem{HG79} Harris, D.E. and Gindlay, J.E. 1979 MNRAS 188, 25

\bibitem{HWDB95} Harris, D. E., Willis, A.G., Dewdney, P.E., and
Batty, J. 1995 MNRAS 273, 785

\bibitem{H98} Henriksen, M. 1998 PASJ 50, 389 

\bibitem{JOS74} Jones, T.W., O'Dell, S.L., and Stein, W.A. 1974 ApJ 188, 353 

\bibitem{KTIIKMOSTA95} Kaneda, H. Tashiro, M., Ikebe, Y., Ishisaki,
Y., Kubo, H., Makshima, K., Ohashi, T., Saito, Y., Tabara, H, and
Takahashi, T. 1995 ApJ 453, L13

\bibitem{M91} McAdam, W. B. 1991 Proc. ASA 9 (2) 255

\bibitem{TKMIIIKTY98} Tashiro, M., Kaneda, H, Makishima, K., Iyomoto,
N., Idesawa, E., Ishisaki, Y., Kotani, T., Takahashi, T., and
Yamashita, A. 1998 ApJ 499, 713


\end{thebibliography}
\end{document}